# Enhanced spin-orbit torques in MnAl/Ta films with improving chemical ordering

K. K. Meng [1, *, †], J. Miao[1], X. G. Xu[1], Y. Wu[1], X. P. Zhao[2], J. H. Zhao[2] and Y. Jiang[1, *, ‡]

[1] *School of Materials Science and Engineering, University of Science and Technology Beijing, Beijing 100083, China*

[2] *State Key Laboratory of Superlattices and Microstructures, Institute of Semiconductors, Chinese Academy of Sciences, Beijing 100083, China*

**Abstract:** We report the enhancement of spin-orbit torques in MnAl/Ta films with improving chemical ordering through annealing. The switching current density is increased due to enhanced saturation magnetization $M_S$ and effective anisotropy field $H_K$ after annealing. Both damplinglike effective field $H_D$ and fieldlike effective field $H_F$ have been increased in the temperature range of 50 to 300 K. $H_D$ varies inversely with $M_S$ in both of the films, while the $H_F$ becomes liner dependent on $1/M_S$ in the annealed film. We infer that the improved chemical ordering has enhanced the interfacial spin transparency and the transmitting of the spin current in MnAl layer.

**Keywords:** Spin polarized transport in metals; Spin–orbit coupling; Spin transport effects

**PACS:** 72.25.Ba; 71.70.Ej; 75.76.+j

[*]Authors to whom correspondence should be addressed:

[†] kkmeng@ustb.edu.cn

[‡] yjiang@ustb.edu.cn

Spin-orbit torques (SOT) are attracting great interest as they can control the magnetization in heavy metal (HM)/ferromagnetic metal (FM) heterostructures with perpendicular magnetic anisotropy (PMA).[1-7] Two classes of spin-orbit coupling (SOC) mechanisms have been considered as candidate sources for SOT: The spin current will be generated in HM layer with strong SOC due to spin Hall effect (SHE), which can diffuse into the adjacent FM layer and exert torques on the moments via spin transfer torque.[1,2] On the other hand, SOT can also arise from the interfacial Rashba effect, for which the accumulated spins at the interface can force the moments to change its direction by direct exchange coupling.[3] Both mechanisms can result in the exertion of dampinglike torques and fieldlike torques on the FM, which can be characterized by equivalent dampinglike effective field $H_D$ and fieldlike effective field $H_F$ respectively.[4,5]

Enhancing these two kinds of torques is the key for practical use of SOT in spintronics devices, and the recent reports have shown that the magnitude and the sign of the torques can greatly depend on the thickness of the FM and HM, the type of FM and HM, the direction of the magnetization in the FM and temperature.[4,5,8-13] Although the origin of the torques is complex and under debate, there are two mechanisms determining the efficiency of SOT. Firstly, enhancing the spin transparency at the interface of HM/FM heterostructures can decrease the spin memory loss and increase the effective torques.[15,16] Pai *et al* have found that a careful engineering of Pt/FM interfaces can improve the spin Hall torque efficiency of Pt-based spintronic devices. The dependence on temperature for both vector components of the spin Hall torque is strongly dependent on the details of the Pt/FM interface.[15] On the other hand, enhancing the absorption of spin currents after they enter the FM layer is considered to be another effective method.[17,18] Qiu *et al* have exploited the large spin absorption at the Ru interface to manipulate the SOT in HM/FM/Ru multilayers.[17] It is found that when the FM thickness is ultrathin enough, the top Ru layer largely boosts the absorption of spin currents into the FM layer and substantially enhances the strength of SOT acting on the FM. In the respect of technique, the above two mechanisms depend on the quality of both the interface and the FM layer in HM/FM heterostructures. In a commonly used stack structure of HM/FM/Oxide with PMA, the magnetic and transport properties of ultrathin FM layer can be modified by annealing at different temperature. However, the PMA originating from interfacial perpendicular anisotropy will degrade after annealing at high temperatures.[6, 19] On the contrary, the PMA originating from bulk rather than interface seems to be stable and also can be improved by thermal treatment.[20-22] Recently, we have systematically investigated the SOT based on $L1_0$-MnGa films with bulk PMA, which seems to be an ideal FM for investigating SOT.[23, 24] On the other hand, $L1_0$-MnAl films with high bulk PMA are also ideal for a systematic examination of SOT, in which the related magnetic and transport behaviors can be controllably varied with thermally tuned chemical ordering.[21, 22]

In this work, we report the SOT in MnAl/Ta bilayers grown on GaAs substrate and compare the results in as deposited and the annealed films. The enhanced $M_S$ and $H_K$ after annealing have resulted in a increased switching current density at room temperature. Considering the thickness of MnAl is very small, the modified PMA is mainly ascribed to the enhanced interfacial anisotropy. By performing adiabatic harmonic Hall voltage measurements in the range from 50 to 300 K, we show that both $H_D$ and $H_F$ have been increased after annealing. Furthermore, $H_D$ appears to vary inversely with $M_S$ in both of the films, while the $H_F$ becomes liner dependent on $1/M_S$ in the annealed film. We infer that firstly the annealing has enhanced the interfacial spin transparency, which in turn promotes the total torques exerted on the MnAl layer. On the other hand, the enhanced chemical ordering also promotes the transmitting of the spin current in MnAl layer to reach the GaAs/MnAl interface. Then the spin scattering at either the MnAl/Ta or GaAs/MnAl interface can promote formation of a net spin accumulation in the FM layer to generate stronger $H_F$.

A 3-nm-thick $L1_0$-MnAl single-crystalline film was grown on a semi-insulating GaAs (001) substrate by molecular-beam epitaxy at substrate temperature $T_S$=250 ℃, in which the Mn/Al atoms ratio is 1.[25] Then, 5-nm-thick Ta film was deposited on it by dc magnetron sputtering. After deposition, the film was annealed at 300 ℃ for 1 hour under an out-of-plane magnetic field of 5 KOe. Photolithography and Ar ion milling were used to pattern Hall bars and a lift-off process was used to form the contact electrodes. The size of all the Hall bars is 10 μm×80 μm. Figure 1(a) shows the schematic of the Hall bar along with the definition of the coordinate system used in this study. We measured the SOT induced magnetization switching by applying a pulsed current with the width 50 μs, and the resistance was measured after a 16 μs delay under an external magnetic field $H_X$ along X direction. We applied a sinusoidal AC current with amplitude 2.1 mA and frequency 239.68 HZ to exert periodic SOT on the magnetization in the temperature range of 50 to 300 K. The first $V_\omega$ and the second $V_{2\omega}$ harmonic anomalous Hall voltages were measured as a function of the magnetic field at the same time using two lock-in amplifier systems. In this work, we have compared the properties between as deposited and annealed films, which have been noted as "as dep" and "an" respectively.

From the shape of the normalized out-of-plane M-H curves measured at 300 K shown in Fig. 1(b), we can find that the coercivity is increased and the PMA is improved after annealing, which agrees with the thermally tuned structural disorders and magnetic properties (e.g., magnetization and PMA).[21] We have also investigated the temperature dependence of saturation magnetization $M_S$ and effective anisotropy field $H_K$ of the two samples as shown in Fig. 1(c) and (d) respectively. The smaller $M_S$ and $H_K$ in the as deposited film indicate the presence of strong disorder effects. In the temperature range of 50 to 300 K, the post annealing has enhanced the values of both $M_S$ and $H_K$, indicating improved chemical ordering of the film.[21] Considering that the thickness of MnAl is very small, the modified interfacial qualities will play a more dominant role on the enhanced PMA, $M_S$ and $H_K$. The current-induced switching measured at 300 K in the as deposited film with an in-plane field of $H_X$=500 Oe is shown in Fig. 1(e), and the SOT with an in-plane field of $H_X$=1000 Oe in the annealed film is shown in Fig. 1(f). Under these two assisted magnetic fields, the magnetization in the two films can be fully switched. As discussed in the previous works, the critical current density $J_C$ originates from either or Rashba effect is linear dependent on both $M_S$ and $H_K$, which can explain the enhanced $J_C$ after annealing.[23]

We have firstly performed the harmonic measurements with sweeping large in-plane magnetic field parallel or perpendicular to the current direction ($H_X$ and $H_Y$). The results in the annealed film were taken as an example as shown in Fig. 2, in which the first harmonic $V_\omega$ and second harmonic Hall voltages $V_{2\omega}$ were plotted against $H_X$ and $H_Y$ at various temperatures. Considering the shapes of $V_\omega$ along X and Y directions are almost the same, we just give the results of $V_\omega$-$H_X$ as shown in Fig. 2(a). With field sweeps along the current flow direction X, the second harmonic signal $V_{2\omega}$ shows a negative peak at a positive magnetic field and a positive peak at a negative magnetic field as shown in Fig. 2(b), which is similar with the results in the HM/FM/Oxide with interfacial induced PMA.[4, 13] The curves of $V_{2\omega}$ along the Y directions show negative peaks around the zero magnetic field as shown in Fig. 2(b), which can be ascribed to the intrinsic magnetic properties of MnAl layer. Furthermore, the $V_{2\omega}$ along the two directions have opposite symmetry with respect to magnetic field: for X direction $V_{2\omega}$ varies antisymmetrically with respect to zero field whereas for Y direction variation is symmetric. As discussed in HM/FM/Oxide structures, these two kinds of torques have odd and even behavior with respect to magnetization reversal, in which the fieldlike torque does not depend on the magnetization direction, whereas dampinglike torque does.[26]

To quantitatively determine the strength of the spin-orbit effective fields in the two samples, the first harmonic $V_{\omega}$ and second harmonic Hall voltages $V_{2\omega}$ against smaller in-plane external field $H_X$ and $H_Y$ were carried out in the temperature range of 50 to 300 K. The harmonic measurements at 300 K are taken as an example as shown in Fig. 3, in which the results are measured with out-of-plane magnetization component $M_Z>0$ and with $M_Z<0$. Before the harmonic measurements shown in Fig. 3, we have applied a large out-of-plane external field to saturate the two samples, which remain saturated after the field is turned off. The dampinglike $H_D$ and fieldlike effective field $H_F$ carried out in the temperature of 50 to 300 K can be calculated by [5]

$$H_{D(F)} = -2 \frac{\partial V_{2\omega}/\partial H_{X(Y)}}{\partial^2 V_\omega/\partial H_{X(Y)}^2} \quad (1)$$

Considering the shapes and values of $V_{\omega}$ along X and Y directions are almost the same, we also just give the results of $V_{\omega}$-$H_X$ of the two samples as shown in Fig. 3(a) and (d). Therefore, according to Equation 1, the value of $\partial^2 V_\omega/\partial H_X^2$ is similar with that of $\partial^2 V_\omega/\partial H_Y^2$, and the values of $H_D$ and $H_F$ mainly depend on the variation of $V_{2\omega}$ with applied $H_X$ and $H_Y$ respectively. It is found that in the as deposited film the $V_{2\omega}$ signal with $H_Y$ applied is very small indicating a smaller fieldlike torque as compared with that in annealed film. The temperature dependence of $H_D$ and $H_F$ in the two samples normalized to the current density of $10^6 A/cm^2$ are summarized in Fig. 4(a) and (b) respectively. The $H_D$ in the two samples are almost the same, but as decreasing the temperature the value in the as deposited film decreases more dramatically. On the other hand, the $H_D$ varies inversely with $M_S$ in both of the films as shown in Fig. 4(c) and (d), which is consistent with relationship of $H_D = \hbar\theta_{SH}|j|/(2|e|M_S t_F)$, where $\theta_{SH}$ is the spin hall angle, j is charge current density, $e$ the charge of an electron and $t_F$ the thickness of MnAl.[27] Using this equation, we have also extracted the effective spin hall angle $\theta_{SH}$=-0.13 of Ta in the as deposited film and the value is -0.17 in annealed film. Using spin-torque ferromagnetic resonance measurements, Zhang *et al* have found that the transparency of the Pt/FM interface to the spin current plays a central role in determining the magnitude of the spin-torque-derived spin Hall angles.[16] They have measured a much larger spin-torque-derived spin Hall angles in Pt/Co compared to Pt/Py bilayers when the interfaces are assumed to be completely transparent. Therefore, the improved MnAl/Ta interface after annealing has enhanced the spin Hall torque and spin transparency, which in turn extort more torques on the MnAl layer. For fieldlike component, a relatively larger enhancement of $H_F$ has been found in the annealed film during the whole temperature range as shown in Fig. 4(b). However, as compared with the linear relationship of $H_D$-$1/M_S$ in the two samples, the $H_F$ only becomes liner dependent on $1/M_S$ in the annealed film. Actually, the origin of the physical origin of the fieldlike effective field in FM/HM bilayers is still debatable. Haney *et al* have developed semiclassical models for electron and spin transport in bilayer nanowires in FM/HM bilayers.[28] They have proved that the dampinglike torque is typically derived from the models describing the bulk SHE and the spin transfer torque, and the fieldlike torque is typically derived from a Rashba model describing interfacial SOC. In the contrary, Ou *et al* have experimentally measured the thickness and temperature dependence of the $H_D$ and $H_F$ in a range of HM/NM/FM/oxide heterostructures.[18] The fieldlike torque in these samples is considered to originate from spin current generated by the SHE in the HM. The spin scattering at either the HM/FM or FM/oxide interface can promote formation of a net spin accumulation in the FM layer that generates $H_F$, provided that a significant portion of the spin current incident on the FM reaches the FM/oxide interface. Recently studies suggest that the $H_F$ can be written in the following form by taking into account the spin Hall current from the HM layer only [29]

$$H_F/j = \frac{\hbar}{2e}\frac{\theta_{SH}}{M_S t_F}(1-\frac{1}{\cosh(d/\lambda_{HM})}) \times \frac{g_i}{(1+g_r)^2+g_i^2} \quad (2)$$

where $g_r = \text{Re}[G_{mix}]\rho\lambda_{HM}\coth(d/\lambda_{HM})$, $g_i = \text{Im}[G_{mix}]\rho\lambda_{HM}\coth(d/\lambda_{HM})$ with $G_{mix}$ the spin-mixing conductance of FM/HM interface, $\rho$ the resistivity of HM, and $\lambda_{HM}$ the spin diffusion length in HM. It is indicated that firstly if the fieldlike effective field is the spin Hall origin, the $H_F$ will becomes linear dependent on $1/M_S$, which is consistent with the relationship of $H_F$-$1/M_S$ in the annealed film. On the other hand, the variation of spin-mixing conductance at the interface will also modify the $H_F$. Therefore, we infer that the enhanced chemical ordering not only modified the spin-mixing conductance at the interface but also promotes the transmitting of the spin current in the more ordered MnAl layer to reach the GaAs/MnAl interface, since the improved ordering has decreased the spin memory loss in the MnAl laye. Then the spin scattering at either the MnAl/Ta or GaAs/MnAl interface can promote formation of a net spin accumulation in the FM layer to generate stronger $H_F$.

Notably, although our work has demonstrated that improving the chemical ordering is an opportunity to tune the SOT based on bulk PMA MnAl, there is a contradiction for this method that the enhanced $M_S$ and $H_K$ have also increased the switching current density, which is not beneficial for the practical use. On the other hand, the Dzyaloshinskii–Moriya interaction (DMI) plays an important role in forming a Néel-type domain wall in thin magnetic multilayers, that can be driven efficiently by the SOT arising from SHE.[30, 31] Yu *et al* have studied the SOT and DMI in the dual-interfaced Co-Ni perpendicular multilayers.[31] The SOT is found to originate mostly from the bulk of a HM, while DMI is more of interfacial origin. A large in-plane assist field is necessary for full switching in the Ta capped sample due to a large negative DMI. In contrast, systems with smaller DMI such as the Cu capped sample only need a moderate assist field to achieve hysteretic reversal. In our work, a larger assisted field is also needed for the annealed film. Meanwhile, the process of the domain wall reversal will also be different in the annealed film due to modified magnetic properties. Therefore, we cannot confirm whether the DMI will also determine the magnetization switching in our systems, especially for the annealed film where the interfacial quality becomes much more dominant. There is still much room for further theoretical and experimental works towards modulating SOT, which is the ultimate goal for technological applications.

In summary, we have compared the magnetic properties and the SOT in the as deposited MnAl/Ta bilayers and the annealed one. The annealing has improved the chemical ordering, resulting in increased $M_S$ and $H_K$. In the annealed film, the switching current density is also increased due to enhanced $M_S$ and $H_K$. However, both $H_D$ and $H_F$ have been increased in the temperature range of 50 to 300 K after annealing. $H_D$ varies inversely with $M_S$ in both of the films, while the $H_F$ only becomes liner dependent on $1/M_S$ in the annealed film. We infer that firstly the annealing has enhanced the interfacial spin transparency, which in turn promotes the total torques exerted on the MnAl layer. On the other hand, the enhanced chemical ordering also promotes the transmitting of the spin current in MnAl layer to reach the GaAs/MnAl interface. Then the spin scattering at either the MnAl/Ta or GaAs/MnAl interface can promote formation of a net spin accumulation in the FM layer to generate stronger $H_F$.

**Acknowledgements**

This work was partially supported by the National Basic Research Program of China (2015CB921502), the National Science Foundation of China (Grant Nos. 61404125, 51371024, 51325101, 51271020).

**Figure Captions**

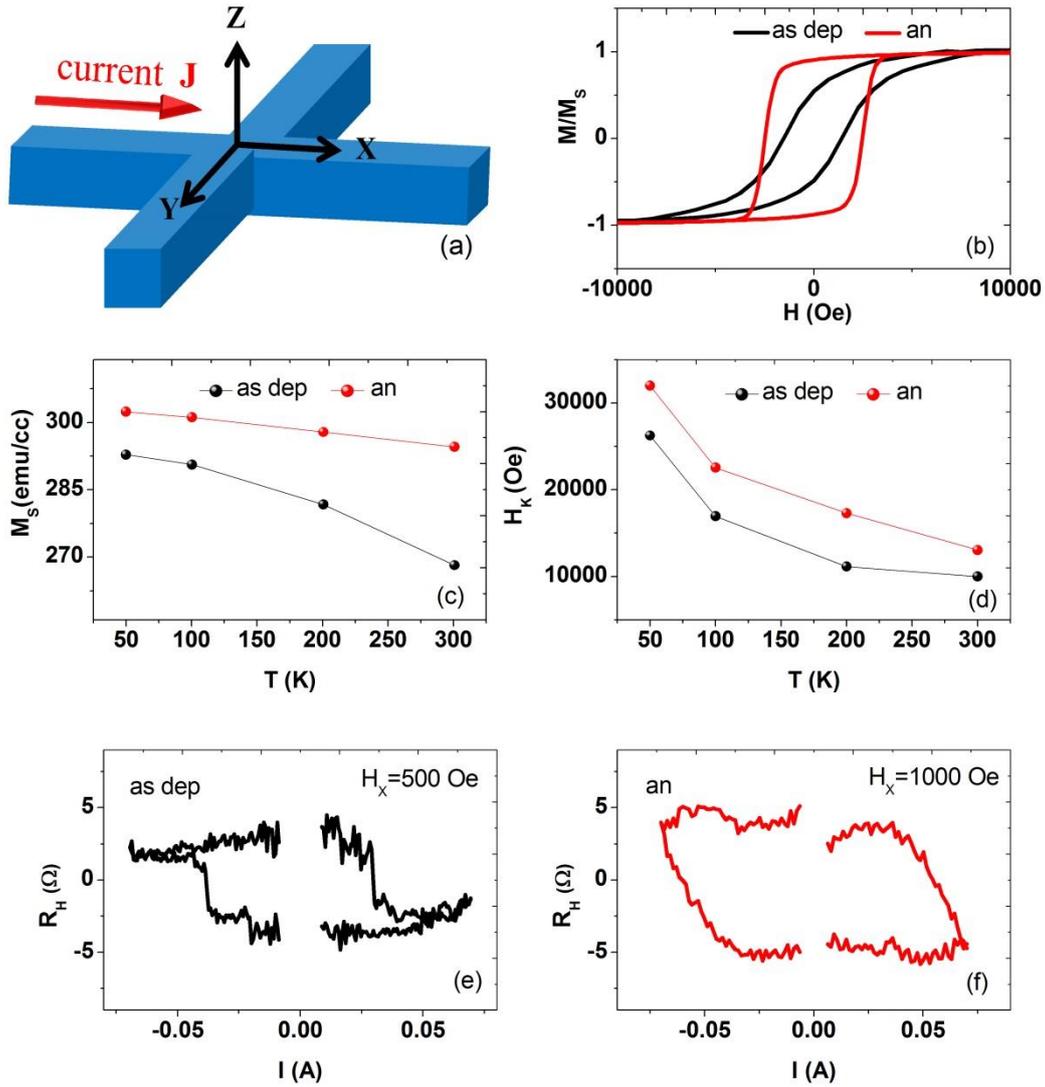

Figure 1. (a) Schematic Hall bar along with the definition of the coordinate system. (b) The out-of-plane M-H curves of the two samples. M is normalized to saturation magnetization $M_S$. The as deposited and annealed films are denoted as "as dep" and "an". (c) Temperature dependence of $M_S$. (d) Temperature dependence of effective anisotropy field $H_K$. $R_H$-I curves of (e) as deposited and (f) annealed film with $H_X$=500 Oe and $H_X$=1000 Oe respectively.

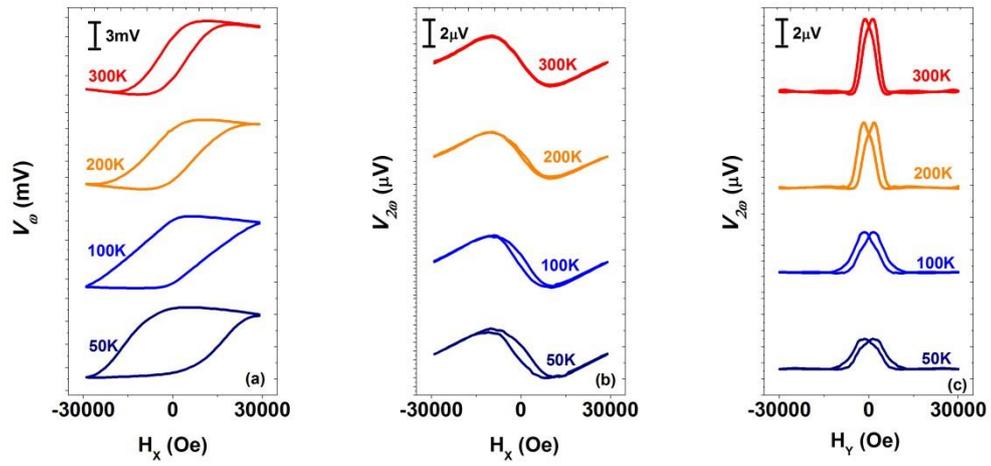

Figure 2. Harmonic measurements in the annealed film with applying large magnetic field from 50 to 300 K. (a) The first harmonic Hall voltages $V_\omega$ plotted against the in-plane external field $H_X$. (b) and (c) show the second harmonic Hall voltages $V_{2\omega}$ plotted against the in-plane external field $H_X$ and $H_Y$ respectively.

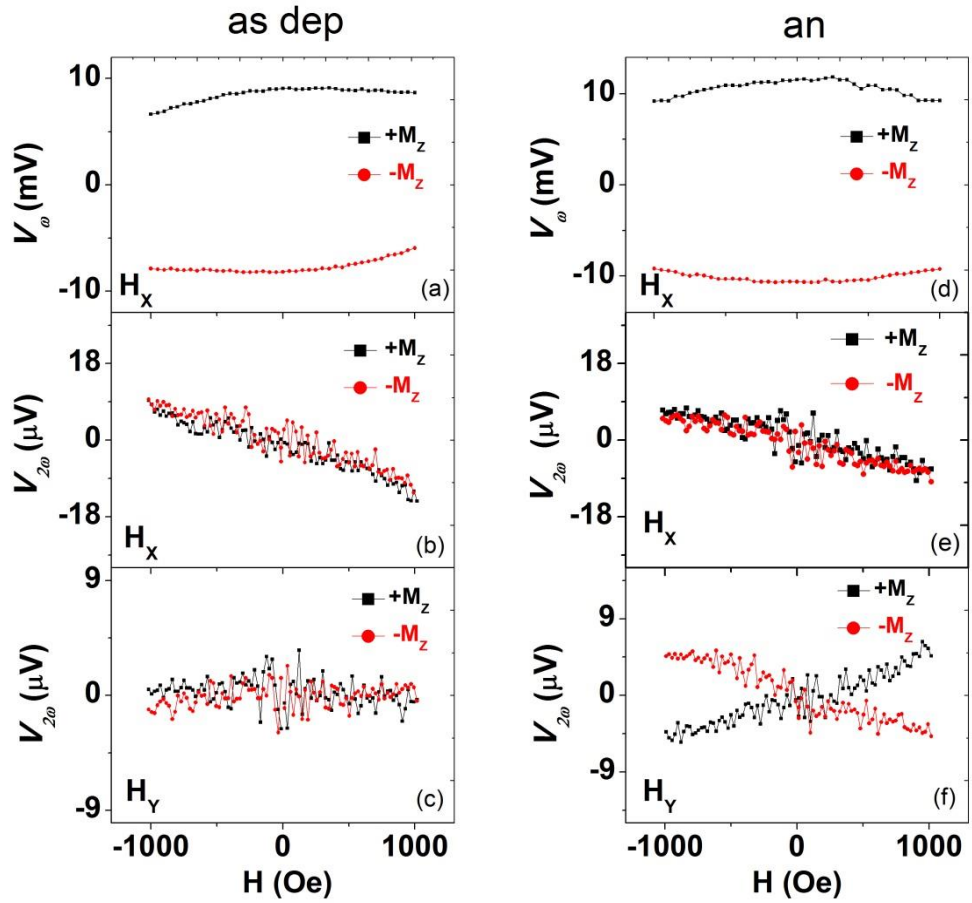

Figure 3. (a)-(c) Harmonic measurements in the as deposited film with applying small field at 300 K. (d)-(f) Harmonic measurements in the annealed film with applying small field at 300 K.

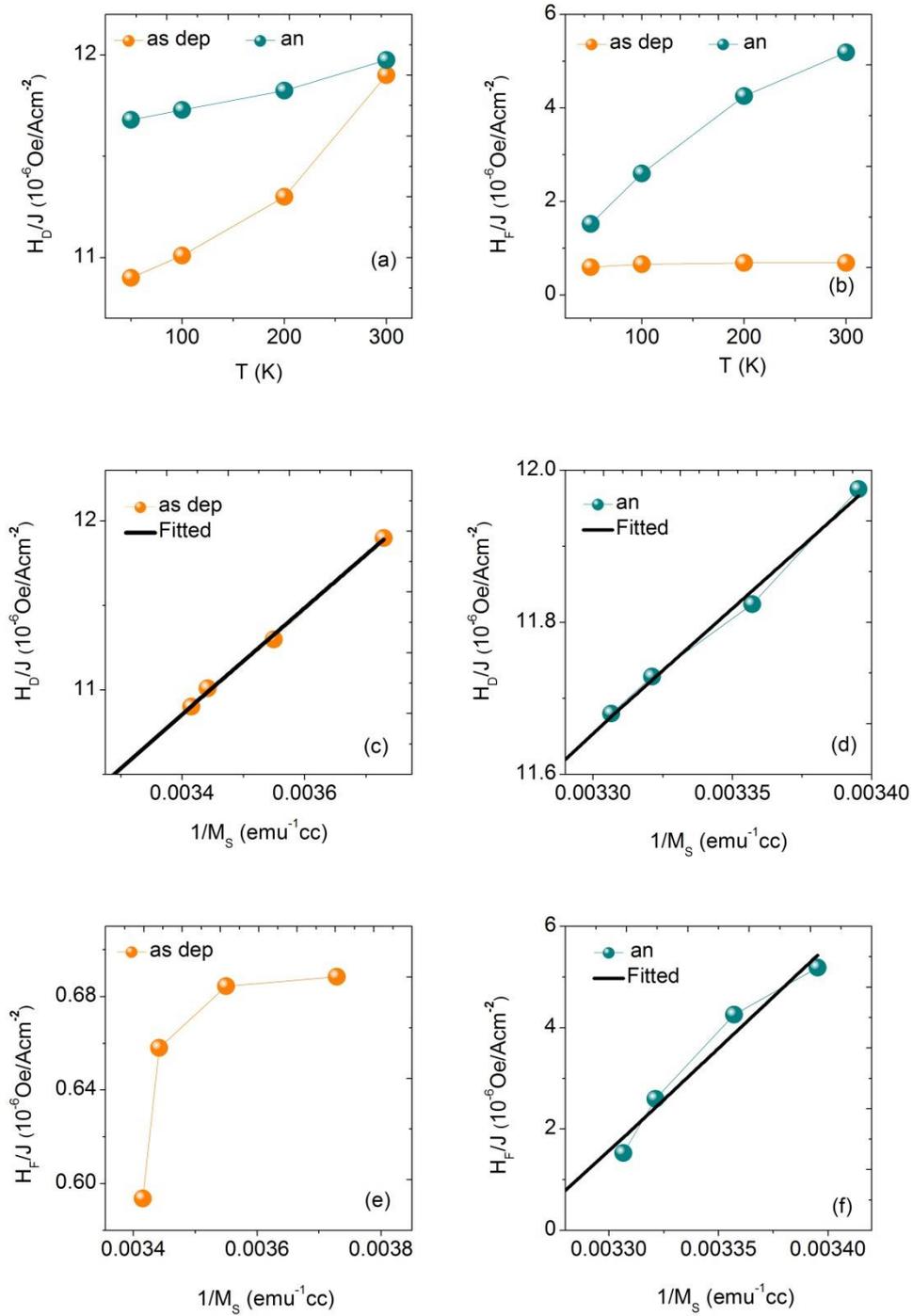

Figure 4. (a) and (b) show temperature dependence of dampinglike effective field $H_D$ and fieldlike effective field $H_F$ in the two samples respectively. (c) and (d) show the $H_D$-$1/M_S$ in the as deposited and annealed films respectively. (e) and (f) show the $H_F$-$1/M_S$ in the as deposited and annealed films respectively.